\documentclass[useAMS,usenatbib,usegraphicx]{mn2e}

\title[Lyman ``bump'' galaxies]
{Lyman ``bump'' galaxies - I. Spectral energy distribution of galaxies
with an escape of nebular Lyman continuum}
\author[A. K. Inoue]{Akio K. Inoue$^{1}$\thanks{E-mail:
akinoue@las.osaka-sandai.ac.jp}\\
$^{1}$College of General Education, Osaka Sangyo University, 
3-1-1, Nakagaito, Daito, Osaka 574-8530, Japan}
\begin{document}

\date{}

\pagerange{\pageref{firstpage}--\pageref{lastpage}} \pubyear{2009}

\maketitle

\label{firstpage}

\begin{abstract}
 It is essential to know galactic emissivity and spectrum of Lyman
 continuum (LyC) in order to understand the cosmic reionization. 
 Here we consider an escape of nebular LyC from galaxies and
 examine the consequent spectral energy distribution. It is usually
 assumed that hydrogen nebular LyC mostly produced by bound-free
 transitions is consumed within photo-ionized nebulae (so-called
 ``on-the-spot'' approximation). However, an escape of the continuum
 should be taken into account if stellar LyC escapes from galaxies
 through ``matter-bounded'' nebulae. We show that the escaping hydrogen
 bound-free LyC makes a strong bump just below the Lyman limit. Such a
 galaxy would be observed as a Lyman ``bump'' galaxy. This bump results
 from the radiation energy re-distribution of stellar LyC by nebulae. 
 The strength of the bump depends on electron temperature in nebulae,
 escape fraction of stellar and nebular LyC, hardness of stellar LyC
 (i.e. metallicity, initial mass function, age and star formation
 history), and IGM attenuation. We can use the bump to find very young
 ($\sim1$ Myr), massive ($\sim100$ $M_\odot$), and extremely metal-poor
 (or metal-free) stellar populations at $z<4$. Because of the bump, 900
 \AA-to-1500 \AA\ luminosity density ratio (per Hz) becomes maximum
 (2--3 times larger than the stellar intrinsic ratio) when about 40\% of
 the stellar LyC is absorbed by nebulae. The total number of escaping LyC
 photons increases due to the escape of nebular LyC but does not exceed
 the stellar intrinsic one. The radiation energy re-distribution by
 nebulae reduces the mean energy of escaping LyC only by $\approx10$\%
 relative to that of stellar LyC. Therefore, the effect of the escape of
 nebular LyC on the reionization process may be small.
\end{abstract}


\section{Introduction}

Understanding the reionization of the intergalactic medium (IGM) is one
of the most important issues in cosmology. The end of the hydrogen
reionization epoch is probably at $z\approx6$ which is inferred from
QSOs \citep[e.g.,][]{bec01,fan06}, Lyman $\alpha$ emitters (LAEs) 
\citep{kas06}, and Gamma-ray bursts \citep{tot06} at the
redshift. Future 21 cm tomography will reveal the ionization history of
hydrogen in detail \citep[e.g.,][]{mad97}. Even after we know the
history, however, the problem of the ionising source may remain still
open. We have not found many objects at $z>6$ so far 
\citep{iye06,sal09,tan09}. Moreover, we do not know the Lyman continuum
(LyC) emissivity of high-$z$ objects.

Galaxies are thought to be more probable source for the hydrogen
reionization than QSOs \citep[e.g.,][]{mad99}. \cite{ino06} found
that the LyC emissivity relative to non-ionising ultraviolet (UV) (or
escape fraction of LyC) should increase by an order of magnitude from
$z<1$ to $z>4$ so as to match the observed UV luminosity density of
galaxies with the hydrogen ionization rate estimated from Lyman $\alpha$
forest. This trend is consistent with a remarkable contrast between
$z\la1$ and $z\approx3$ in direct observations of LyC from galaxies;
there is no detection of LyC at $z\la1$ \citep{lei95,hur97,deh01,sia07,cow09}, 
but a marginal exception \citep{ber06,gri07}, whereas there are direct
detections of LyC from Lyman break galaxies (LBGs) and LAEs at
$z\approx3$ \citep{ste01,sha06,iwa09}. This evolving escape fraction is
also found in a simulation of galaxy formation with the radiative
transfer \citep{raz06,raz07}. 

The latest observations with the Subaru telescope have found LAEs
emitting extremely strong LyC \citep{iwa09}. Indeed, some of them are
brighter in LyC ($\approx900$ \AA) than in UV ($\approx1500$ \AA). Such
extreme ``blue'' colours cannot be explained by standard stellar
populations with a Salpeter initial mass function \citep{iwa09}.
If we estimated the escape fraction for them, we would obtain a value
larger than unity. This shows our poor knowledge of LyC emissivity
and spectrum of galaxies. We should consider the LyC spectrum of
star-forming galaxies more carefully.

This paper examines effects of nebular emission on the spectral energy
distribution (SED) of galaxies. Such studies have been made so far by
several authors \citep{sta96,moy01,zac01,sch02,zac08,sch09}. However, we
take into account, for the first time, an escape of nebular LyC from
photo-ionized nebulae and galaxies. The nebular LyC is usually assumed
to be absorbed within the nebulae, the so-called ``on-the-spot''
approximation \citep{ost06}. This is a very good treatment for
``photon-bounded'' nebulae from which no LyC escapes. If stellar LyC
escapes from ``matter-bounded'' nebulae and escapes from galaxies as
observed at $z\approx3$, however, nebular LyC may also escape. We
discuss this latter case here.

The rest of this paper consists of four sections. In Section 2, we
describe modelling procedures: stellar population models, calculations
of nebular emission, and IGM attenuation. The predicted spectral energy
distributions, 900 \AA-to-1500 \AA\ luminosity density ratios, and other
results are presented in Section 3. In Section 4, we discuss validity of
assumptions made in modelling and a few implications from the
results. The final section is devoted to the conclusion.

\section{Model}

\subsection{Stellar emission}

We use population synthesis models taken from literature. 
Parameters characterising SEDs of galaxies are metallicity and
initial mass function (IMF) as well as star formation history and age. 
Motivated by the discovery of LAEs emitting very strong LyC
\citep{iwa09}, we consider stellar populations emitting LyC as strongly
as possible. Thus, we consider very young age of 1 Myr after an
instantaneous burst, except for \S3.2 where we also consider a
continuous star formation with a longer age. The metallicity assumed
here is $Z=0.0004$ which is 1/50 of the Solar metallicity and the lowest
one available in the code Starburst99 version 5.1
\citep{lei99}.\footnote{The stellar track assumed is a Padova track with
AGB stars. The atmosphere model assumed is "PAULDRACH/HILLIER" as the
recommendation of the code. This choice of the atmosphere model may
affect the stellar LyC emissivity \citep{sch03}.} Then, we consider two
IMF slopes of $-2.35$ \citep{sal55} and $-0.1$ (extremely top-heavy). We
call this two models A and B, respectively. The mass range is assumed to
be 1--100 $M_\odot$ for both models. In addition, we consider two more
cases: $Z=10^{-5}$ (extremely metal-poor [EMP] stars; \citealt{bee05})
and $Z=0$ (metal-free or Population-III stars) with the Salpeter IMF
slope and 50--500 $M_\odot$. These models are taken from \cite{sch03}
and called C and D, respectively. Table 1 is a summary of the stellar
population models considered in this paper.

\begin{table}
 \caption[]{Models of stellar population and SED.}
 \setlength{\tabcolsep}{3pt}
 \footnotesize
 \begin{minipage}{\linewidth}
  \begin{tabular}{lccccc}
   \hline
   Model & $Z$ & $p$ & $M_{\rm up}$ & $M_{\rm low}$ & SED reference\\
   \hline
   A & $0.0004$ & $-2.35$ & $100 M_\odot$ & $1 M_\odot$ & SB99 (v.5.1)\\
   B & $0.0004$ & $-0.1$ & $100 M_\odot$ & $1 M_\odot$ & SB99 (v.5.1)\\
   C & $10^{-5}$ & $-2.35$ & $500 M_\odot$ & $50 M_\odot$ 
		   & Schaerer (2003)\\
   D & $0$ & $-2.35$ & $500 M_\odot$ & $50 M_\odot$ & Schaerer (2003)\\
   \hline
  \end{tabular}
 \end{minipage}
\end{table}%

We have to note that modelling of stellar LyC is still developing. 
For example, stellar rotation, which is not taken into account here, 
may increase the LyC emissivity of stars with an age between 3--10 Myr
by a factor of about 2 \citep{vaz07}. Thus, stellar LyC models have such
an amount of uncertainty.

\subsection{Nebular emission}

\subsubsection{Structure of the ISM}

\begin{figure}
 \begin{center}
  \includegraphics[width=7cm]{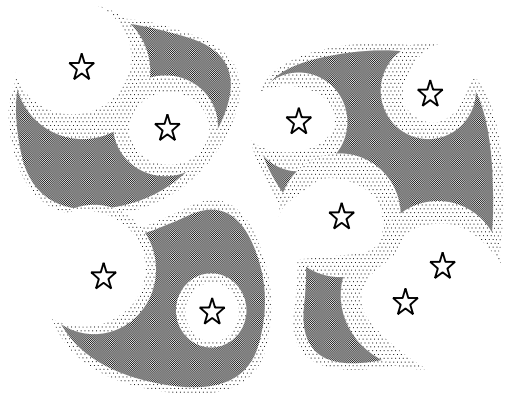}
 \end{center}
 \caption{Schematic picture of clumpy ISM. Dense regions which remain
 neutral against the ionization by the stellar radiation are shown by
 the thick shades and ionized nebulae formed at the surface of the
 dense regions are shown by the thin shades. Other space is filled
 with diffuse ionized gas which has negligible opacity for Lyman
 continuum.}
\end{figure}

The interstellar medium (ISM) in galaxies is clumpy. There should be
dense clumps which remain neutral against the ionising radiation from
stars. Considering a typical density of neutral hydrogen ($\sim10^3$ 
cm$^{-3}$) and a typical size ($\sim5$ pc) found in dark clouds 
\citep{mye78}, we find $10^5$ as the optical depth at the Lyman limit
and $10^2$ even at 100 \AA. Thus, all the LyC photons along a ray
from an ionising star are absorbed if the ray intersects a dense
clump. If the inter-clump medium is diffuse and ionized highly enough to
have a negligible opacity for LyC\footnote{If not, we could not observe
strong LyC from galaxies. In addition, shock ionization by multiple
supernovae would contribute to keep low neutral fraction in the
inter-clump medium as shown by \cite{yaj09}.}, the escape fraction of
the photons is determined by the covering fraction of the clumps around
an ionising star. Thus, it becomes independent of the
wavelength. Ionized regions are formed at the surface of the neutral
clumps. The nebular emission from the regions includes LyC. The escape
fraction of the nebular LyC is assumed to be the same as that of the
stellar LyC for simplicity. This may be good in average sense when stars
and clouds are well mixed. Figure~1 is a schematic picture of such a
clumpy medium.

\subsubsection{Luminosity density}

Let us denote the escape fraction of LyC as $f_{\rm esc}$. This is the
number fraction of LyC photons which escape from a galaxy relative to
the photons produced in the galaxy. As described above, we omit the
frequency dependence of $f_{\rm esc}$. However, we will see the effect
in Section 4.1.

The ionization equilibrium of hydrogen in photo-ionized nebulae with
$f_{\rm esc}$ is
\begin{equation}
 Q_* (1-f_{\rm esc}) + Q_{\rm neb} (1-f_{\rm esc})
  = \int n_{\rm p} n_{\rm e} \alpha_{\rm A}(T_{\rm e}) dV\,,
\end{equation}
where $Q_*$ and $Q_{\rm neb}$ are the production rates of LyC photons by
stars and by nebulae, respectively, $n_{\rm p}$ and $n_{\rm e}$ are
proton and electron number densities, respectively, and $\alpha_{\rm A}$
is the recombination rate to all states. The integral is performed over
the entire volume of the nebulae, $V$. The recombination rate depends on
the local electron temperature in the nebulae, $T_{\rm e}$ \citep{ost06}. 
If we assume $T_{\rm e}$ to be uniform in the nebulae, we can write 
$\int n_{\rm p}n_{\rm e}\alpha_{\rm A}dV=\alpha_{\rm A}\int n_{\rm p} n_{\rm e}dV$.
The production rate of nebular LyC photons can be expressed as 
$Q_{\rm neb}=\alpha_1 \int n_{\rm p} n_{\rm e}dV$ (for a uniform 
$T_{\rm e}$) because LyC photons are produced by the recombination
to the ground state (its rate is $\alpha_1$), except for the free-free
contribution which is negligible in photo-ionized nebulae. Therefore,
equation (1) is reduced to  
\begin{equation}
 \int n_{\rm p} n_{\rm e} dV 
  = \frac{Q_* (1-f_{\rm esc})}
  {\alpha_{\rm B}(T_{\rm e}) + \alpha_1(T_{\rm e})f_{\rm esc}}\,.  
\end{equation}
Note that the ‘Case B’ recombination rate is 
$\alpha_{\rm B}=\alpha_{\rm A}-\alpha_1$ \citep{ost06}. 

The luminosity density of the nebular emission at frequency $\nu$ is
\begin{equation}
 L^{\rm neb}_\nu = 
  \int \gamma^{\rm neb}_\nu(T_{\rm e}) n_{\rm p} n_{\rm e} dV\,,
\end{equation}
where $\gamma^{\rm neb}_\nu$ is the nebular emission coefficient which
does not only depend on the temperature $T_{\rm e}$ but also on the
density. However, we can omit the density dependence here (see \S2.2.3).
From equations (2) and (3), we obtain
\begin{equation}
 L^{\rm neb}_\nu = \gamma^{\rm neb}_\nu(T_{\rm e}) 
  \frac{Q_* (1-f_{\rm esc})}{\alpha_B(T_{\rm e}) 
  + \alpha_1(T_{\rm e})f_{\rm esc}}\,,
\end{equation}
for a uniform $T_{\rm e}$. Note that equation (4) is equivalent to the
standard treatment of nebular emission, for example, equation (2) in
\cite{sch02}, if $f_{\rm esc}=0$. The evaluation of equation (4) needs
recombination rates. We obtain the rates from an interpolation of
Table~3 in \cite{fer80}.

\subsubsection{Emission coefficients}

\begin{figure}
 \begin{center}
  \includegraphics[width=7cm]{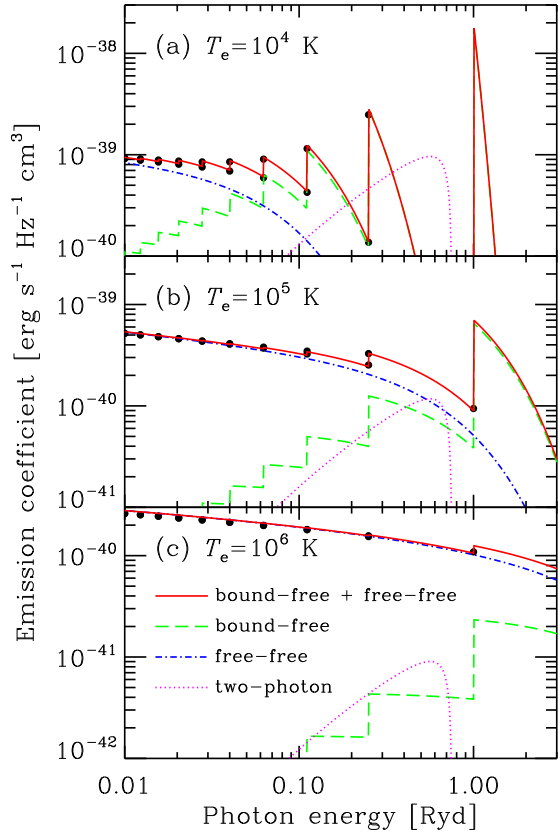}
 \end{center}
 \caption{Hydrogen nebular emission coefficients for (a) electron
 temperature $T_{\rm e}=10^4$ K, (b) $10^5$ K, and (c) $10^6$ K. 
 The bound-free (dashed), free-free (dot-dashed), and two-photon
 (dotted) coefficients are shown. The solid line in each panel is the
 sum of the bound-free and the free-free and it should be compared with
 the calculations by Ferland~(1980) which are shown by small filled
 circles.}
\end{figure}

\begin{figure*}
 \begin{center}
  \includegraphics[width=15cm]{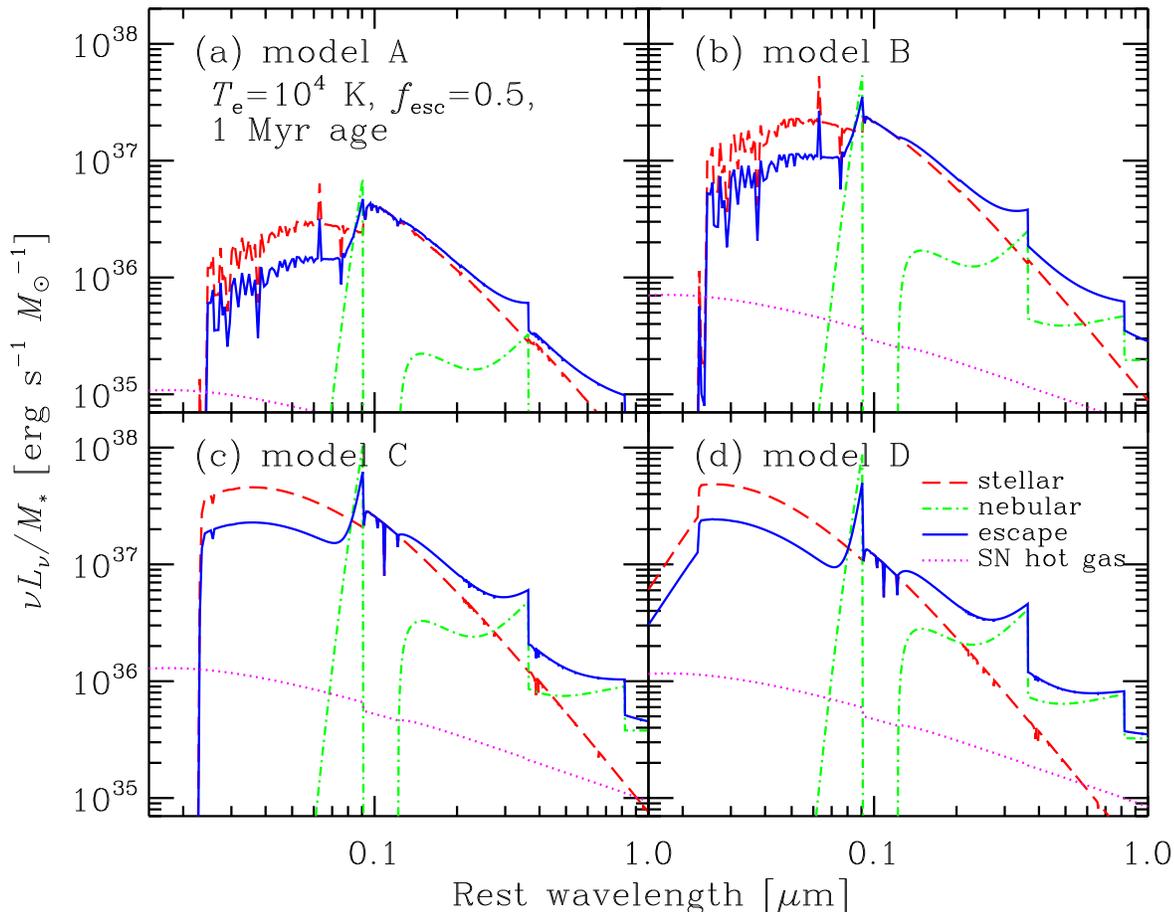}
 \end{center}
 \caption{Spectral energy distributions (SEDs) with escaping nebular Lyman
 continuum: (a)--(d) for models A--D in Table 1. The assumed parameters
 are shown in panel (a): electron temperature $T_{\rm e}=10^4$ K in
 nebulae, escape fraction $f_{\rm esc}=0.5$, and 1 Myr after an
 instantaneous burst. The stellar (dashed), nebular (dot-dashed),
 escaping stellar+nebular (solid), and hot gas (dotted; $T_{\rm e}=10^6$
 K and 5\% of the bolometric luminosity) SEDs are shown in each panel. 
 The vertical axis is normalized by the mass of the star cluster.} 
\end{figure*}

The nebular emission coefficient is the sum of three emission processes
of bound-free, free-free, and two-photon as
\begin{equation}
 \gamma^{\rm neb}_\nu(T_{\rm e}) = \gamma^{\rm bf}_\nu(T_{\rm e}) 
  + \gamma^{\rm ff}_\nu(T_{\rm e}) + \gamma^{\rm 2q}_\nu(T_{\rm e})\,.
\end{equation}
We consider, for simplicity, nebulae composed of hydrogen only since the
effect of helium is small as shown later (\S4.2).
The free-free and bound-free emission coefficients are calculated by
equations (3) and (9) of \cite{mew86}. The two-photon emission
coefficient is calculated as \citep{ost06} 
\begin{equation}
 \gamma^{\rm 2q}_\nu = 2hyP(y)X_{\rm 2q}\alpha_{\rm B}\,,
\end{equation}
where $h$ is the Planck constant, $y$ is the frequency normalized by
that of Ly$\alpha$, $P(y)$ is the normalized spectral shape whose
approximated functional form is given by \cite{nus84}, and $X_{\rm 2q}$
is the probability of being in 2 $^2$S state after one Case B
recombination. We adopt $X_{\rm 2q}=0.32$ \citep{spi51}. The two-photon
emissivity depends on density if nebulae have a high density, but only
the small density limit ($<10^4$ cm$^{-3}$; \citealt{ost06}) would be
enough for this paper. We also note that the Case B approximation, which
is suitable for nebulae optically thick for Lyman series lines
\citep{ost06}, does not conflict with our assumption of nebulae
optically thin for LyC because the absorption cross sections for Lyman
lines are a few orders of magnitude larger than that for LyC.

Figure~2 shows the nebular emission coefficients for $T_{\rm e}=10^4$, 
$10^5$, and $10^6$ K: bound-free (dashed lines), free-free
(dot-dashed lines), and two-photon (dotted lines). We have confirmed
that sums of the bound-free and free-free coefficients (solid lines)
agree with the values presented in Table~1 of \cite{fer80} for
$\lambda>912$ \AA\ (small filled circles) within about 10\% difference.

\subsubsection{Hot gas heated by supernovae and stellar winds}

In addition to photo-ionized nebulae, shock-ionized nebulae produced by
multiple supernovae and stellar winds may contribute to LyC emission. 
According to \cite{lei99}, the mechanical luminosity produced by
supernovae and stellar winds is $\sim1$\% of the bolometric luminosity
of stars and well less than 10\% of it. We assume here 5\%, a somewhat
large value, to show that this component is less important. We also
assume the temperature to be $10^6$ K for this component.

\subsection{IGM attenuation}


Radiation with a wavelength shorter than Ly$\alpha$ in the rest-frame of
the source is absorbed by neutral hydrogen remained in the IGM. We use a
Monte Carlo simulation of the IGM attenuation by \cite{ino08}. This
simulation is based on an empirical distribution function of the column
density, the number density, and the Doppler parameter of absorbers in
the IGM which is derived from the latest observational statistics. The
predicted Ly$\alpha$ decrements agree with the observations in the full
range of $z=0$--6 excellently.

%

\section{Result}

\subsection{Spectral energy distribution}

\begin{figure}
 \begin{center}
  \includegraphics[width=7cm]{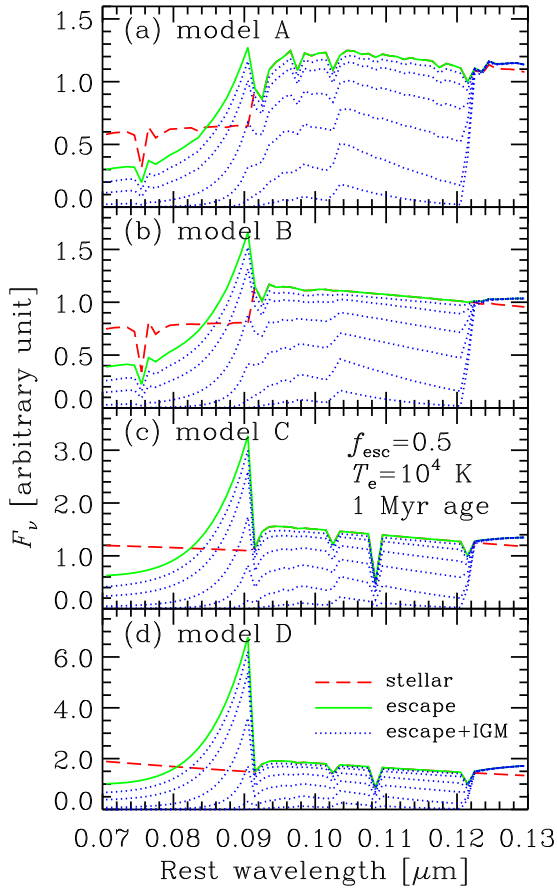}
 \end{center}
 \caption{Close-up spectra around the Lyman limit: (a)--(d) for models
 A--D in Table 1. The assumed parameters are shown in panel (c):
 electron temperature $T_{\rm e}=10^4$ K in nebulae, escape fraction 
 $f_{\rm esc}=0.5$, and 1 Myr after an instantaneous burst. The stellar
 (dashed) and escaping stellar+nebular (solid) spectra are the same as
 Figure~3. The dotted lines show spectra affected by IGM mean
 attenuation. Redshift of the galaxy is $z=1$ to 6 from top to bottom in
 each panel. The vertical axis is normalized by the flux density at
 Ly$\alpha$.}
\end{figure}

\begin{figure}
 \begin{center}
  \includegraphics[width=7cm]{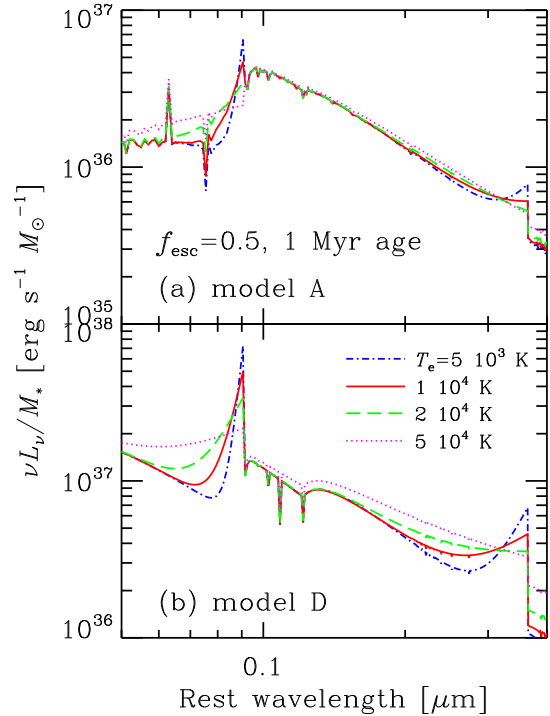}
 \end{center}
 \caption{Effect of electron temperature $T_{\rm e}$ on the escaping
 SEDs: (a) for the model A and (b) for the model D. The assumed escape
 fraction is $f_{\rm esc}=0.5$ and the age is 1 Myr after an
 instantaneous burst. Four cases are shown in each panel: 
 $T_{\rm e}=5\times10^3$ K (dot-dashed), $1\times10^4$ K (solid),
 $2\times10^4$ K (dashed), and $5\times10^4$ K (dotted).}
\end{figure}

First, we show the resultant SEDs without IGM attenuation in Figure~3.
Panels (a)--(d) correspond to models A--D in Table~1. The assumed
quantities are $T_{\rm e}=10^4$ K, $f_{\rm esc}=0.5$, and age of 1 Myr
after an instantaneous burst. The dashed lines are the stellar SEDs. 
We can see that LyC becomes harder from the model A to D 
(see also Table~3). The dot-dashed lines are nebular emissions produced
in photo-ionized nebulae. We can see bound-free Lyman, Balmer, and
Paschen continua and two-photon continuum. Free-free emission is also
taken into account but negligible for this case. The solid lines are
escaping stellar+nebular emissions: sum of the two emissions but reduced
by a factor of $f_{\rm esc}$ in LyC. We can see a significant
contribution of the nebular emission. In particular, the bound-free LyC
makes a bump just below the Lyman limit for models B--D. Even for the
model A, the escaping LyC is a factor of 2 larger than the stellar one
at $\approx900$ \AA. 

The significant contribution of the nebular LyC is due to the energy
re-distribution by nebulae. When a LyC photon ionizes a hydrogen atom,
the photon energy moves to a photo-electron. The energy is thermalized
in nebulae collisionally. Then, an electron recombines with a proton. If
the recombination is to the ground state, we get a photon with an energy
of 13.6 eV + kinetic energy of the electron. If $T_{\rm e}\sim10^4$
K, the kinetic energy is an order of 1 eV. Thus, the photon wavelength
is not far from the Lyman limit. Moreover, the recombination probability
increases for an electron with a lower kinetic energy. Therefore,
nebular LyC shows a peak at the Lyman limit. This is already found in
the emission coefficient shown in Figure~2. 

The dotted lines in Figure~3 are emissions from hot gas produced by
supernovae and stellar winds. We can conclude that this component is less
important On the other hand, \cite{yaj09} have shown an importance of
shock ionization for a large escape fraction; shock ionization in
addition to photo-ionization keeps the content of neutral hydrogen lower
and suppresses the LyC opacity through the ISM. This is not conflict with
our argument that LyC emitted by shocked gas itself is negligible.

Figure~4 shows model spectra affected by IGM attenuation (dotted
lines). The assumed parameters are the same as Figure~3, and thus, the
stellar (dashed) and escaping stellar+nebular spectra (solid) are the
same as Figure~3. The spectra are normalized by the flux density at
Ly$\alpha$ and the unit of the vertical axis is arbitrary if it is for
flux density (per Hz). We can see a prominent ``bump'' just below the
Lyman limit due to the nebular bound-free LyC. Neutral hydrogen remained
in the IGM absorbs radiation below Ly$\alpha$. We apply IGM mean
attenuation by \cite{ino08} as described in \S2.3.1. Each dotted line
corresponds to different redshift at which the object lies: $z=1$, 2,
..., 6 from top to bottom. We find that the Lyman limit bump is visible
up to $z=3$ in this figure. Spectroscopic or narrowband observations
tracing this wavelength range for $z\la3$ is quite interesting to
confirm the reality of the model. If the object lies along a sight-line
less opaque than the average, we may find the bump from $z\approx4$ but
not from $z>5$ (see Figs.~8 and 11 in \citealt{ino08}).

Figure~5 shows the effect of the assumed electron temperature 
$T_{\rm e}$ on the escaping stellar+nebular SEDs: 
$T_{\rm e}=5\times10^3$ K (dot-dashed lines), $1\times10^4$ K (solid), 
$2\times10^4$ K (dashed), and $5\times10^4$ K (dotted). Other parameters
($f_{\rm esc}$ and age) are the same as Figure~3. We show the models A
and D only but the models B and C are qualitatively similar. We find that 
the shape of the Lyman limit bump becomes more peaked for lower $T_{\rm e}$.
This is because the number of electrons with larger kinetic energy
is smaller for lower $T_{\rm e}$ in the Maxwell-Boltzmann distribution,
thus, the photon energy emitted by recombination becomes closer to the
Lyman limit for lower $T_{\rm e}$. The same thing is true for the
behavior around the Balmer limit.

\subsection{900 \AA-to-1500 \AA\ luminosity density ratio}

When estimating the escape fraction of LyC from direct observations, we
often have to assume an intrinsic stellar ratio of LyC to non-ionizing
UV luminosity densities \citep[see \S4 of][]{ino05}.\footnote{In
literature, 1500 \AA-to-900 \AA\ ratio is more popular, but we here take
the inverse to avoid the divergence when $f_{\rm esc}\to0$.} For
example, \cite{ste01} assumed $L_{\nu900}/L_{\nu1500}=0.33$ for their 
$z\sim3$ LBGs and \cite{sia07} argued that 
$L_{\nu900}/L_{\nu1500}=0.17$ is better for their
$z\sim1$ galaxies based on SED fitting. However, escaping
stellar+nebular SEDs shown in Figure~3 are completely different from the
stellar intrinsic ones because of the Lyman limit bump by the nebular
bound-free LyC. Here, we present the 900 \AA-to-1500 \AA\ luminosity
density ratio of escaping stellar+nebular spectra.

\begin{figure}
 \begin{center}
  \includegraphics[width=7cm]{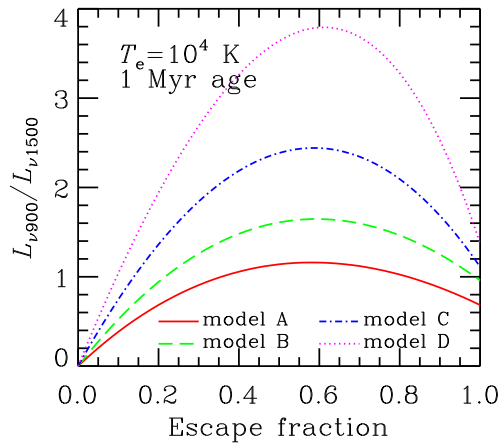}
 \end{center}
 \caption{Luminosity density ratio (per Hz) of Lyman continuum (900 \AA)
 to non-ionizing ultraviolet (1500 \AA) as a function of escape
 fraction. The assumed parameters are electron temperature 
 $T_{\rm e}=10^4$ K and 1 Myr after an instantaneous burst. Models A--D
 in Table 1 are shown with each line style.}
\end{figure}

Figure~6 shows the 900 \AA-to-1500 \AA\ luminosity density ratio as a
function of escape fraction $f_{\rm esc}$ for the case with 
$T_{\rm e}=10^4$ K and 1 Myr after an instantaneous burst. Four models
in Table~1 are shown by each line style: solid for model A, dashed for
model B, dot-dashed for model C, and dotted for model D. At 
$f_{\rm esc}=1$, there is no nebula, and we see the stellar intrinsic
ratio for each model. As $f_{\rm esc}$ decreases towards 0, the ratio
increases, presents a peak, decreases, and reaches 0. The peak appears
at $f_{\rm esc}=0.58$--0.62 and the peak ratio is a factor of 1.7--2.8
larger than the stellar ratio. Note that the 900 \AA-to-1500 \AA\
luminosity density ratio becomes maximum not when 100\% of the stellar
LyC escapes but when 40\% of it is absorbed by nebulae. This is the
Lyman bump effect due to the nebular bound-free emission.  

%

\begin{figure}
 \begin{center}
  \includegraphics[width=7cm]{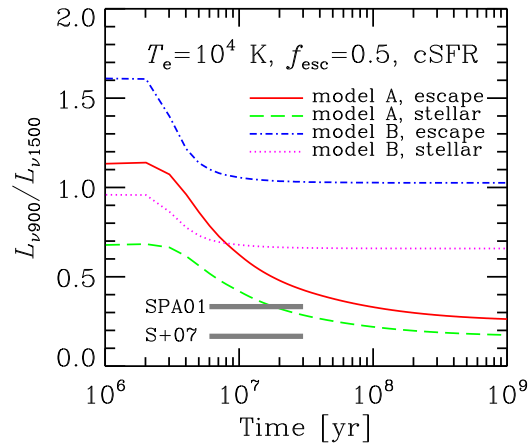}
 \end{center}
 \caption{900 \AA-to-1500 \AA\ luminosity density ratio (per Hz) as a
 function of the duration of constant star-formation for models A and
 B. For model A, stellar ratio is the dashed line and escaping
 stellar+nebular ratio is the solid line. For model B, stellar ratio is
 the dotted line and escaping stellar+nebular ratio is the dot-dashed
 line. The electron temperature $T_{\rm e}=10^4$ K and escape fraction
 $f_{\rm esc}=0.5$ are assumed. Two gray thick marks indicate the values
 assumed in Steidel et al.~(2001, SPA01) and Siana et al.~(2007, S+07).}
\end{figure}

Figure~7 shows the effect of the age on 900 \AA-to-1500 \AA\ luminosity
density ratio. We again assume $T_{\rm e}=10^4$ K and $f_{\rm esc}=0.5$
and also assume a constant star formation for this analysis. We only
show the results of the models A and B. Overall feature found in
Figure~7 is that the ratio decreases after the first a few Myr and
reaches an asymptotic value. This is because 900 \AA\ radiation comes
only from very massive stars whose life-time is less than a few Myr but
1500 \AA\ radiation comes from not only such massive stars but also
relatively long-lived intermediate mass stars. Thus, the 1500 \AA\
luminosity density increases until the intermediate mass stars have
turnover. The contribution of such intermediate mass stars to 1500 \AA\
radiation depends on the IMF. For the model B, the contribution is
small, and then, the luminosity density ratio reaches the asymptotic
value earlier than the model A. For models C and D, the ratio is almost
independent of the age because there is no intermediate mass stars in
these models.

Let us compare the luminosity density ratio predicted here with those
assumed in literature which are shown by gray thick marks in Figure~7.
The ratio proposed by \cite{ste01} corresponds to the stellar intrinsic
ratio (dashed line) for 10--20 Myr age of the model A but to the
escaping stellar+nebular ratio (solid) for 100 Myr age of the model.
The ratio proposed by \cite{sia07} corresponds to the stellar intrinsic
ratio for $>100$ Myr age of the model A. On the other hand, we expect
much larger ratios for younger age: $L_{\nu900}/L_{\nu1500}=1.1$ for the
model A with the nebular contribution if age less than a few Myr. This
ratio is a factor of 3 larger than \cite{ste01} and a factor of 6 larger
than \cite{sia07}. If we change the IMF (model B) or metallicity and mass
range (models C and D), even larger ratios are expected (see Figure~6).

\subsection{Total number of escaping LyC photons}

We have seen that the 900 \AA\ luminosity density increases due to the
nebular bound-free emission. This is the radiation energy 
re-distribution by nebulae. How about the total number of photons in
escaping LyC? The LyC photon escape rate is given by 
\begin{equation}
 Q_{\rm esc} = (Q_* + Q_{\rm neb}) f_{\rm esc}\,, 
\end{equation}
with the escape fraction of stellar and nebular LyC, $f_{\rm esc}$, and 
the production rates of LyC photons by stars, $Q_*$, and by nebulae,
$Q_{\rm neb}$. From the ionization equilibrium in equation (1) combined
with equation (2), equation (7) is reduced to 
\begin{equation}
 \frac{Q_{\rm esc}}{Q_*} = \frac{\alpha_{\rm A} f_{\rm esc}}
  {\alpha_{\rm B} + \alpha_1 f_{\rm esc}}\,.
\end{equation}
The $Q_{\rm esc}/Q_*$ can be called ``effective'' escape
fraction and depends on $f_{\rm esc}$ and $T_{\rm e}$ through
$\alpha$s. Note that the escape fraction $f_{\rm esc}$ is the fraction
of stellar and nebular LyC photons which escape from a galaxy relative
to all (stellar+nebular) LyC photons produced in the galaxy (see
eq.~[7]). On the other hand, the ``effective'' escape fraction is the
fraction of escaping stellar+nebular LyC photons relative to the photons
produced by only stars.

\begin{figure}
 \begin{center}
  \includegraphics[width=7cm]{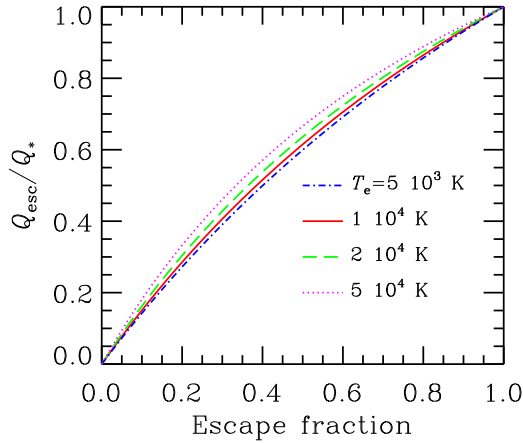}
 \end{center}
 \caption{Ratio of the stellar+nebular escape rate to the stellar
 production rate of LyC photons, or ``effective'' escape fraction, as a
 function of escape fraction. Four lines show different electron
 temperature $T_{\rm e}$: $5\times10^3$ K (dot-dashed), $1\times10^4$ K 
 (solid), $2\times10^4$ K (dashed), and $5\times10^4$ K (dotted).}
\end{figure}

Figure~8 shows the number fraction of LyC photons escaped relative to
those produced by stars, or ``effective'' escape fraction, as a function
of $f_{\rm esc}$ given by equation (8). The escaping number fraction is
always larger than $f_{\rm esc}$ but never exceeds unity. That is, the
total number of escaping LyC photons is always smaller than the number
of LyC photons produced by stars. Nebular LyC compensates only a
fraction of stellar LyC photons which absorbed by nebulae. Since the
nebular LyC has a strong peak at the Lyman limit, however, the
luminosity density near the limit is significantly enhanced as discussed
in \S3.2.

\subsection{Mean energy of escaping Lyman continuum}

Because of modification in spectra by nebular emissions, the mean 
energy of LyC photons escaping from galaxies is changed from that in the
stellar spectrum. Let us define the mean energy of escaping LyC photons
as 
\begin{equation}
 \langle \epsilon_{\rm esc} \rangle \equiv 
  \frac{L_{\rm esc}^{\rm LyC}}{Q_{\rm esc}}
  = \frac{L_*^{\rm LyC} + L_{\rm neb}^{\rm LyC}}{Q_* + Q_{\rm neb}}\,,
\end{equation}
where $L^{\rm LyC}$s are frequency integrated LyC luminosities of
escaping, stellar, or nebular radiations. The equal is true for
frequency independent $f_{\rm esc}$. The $Q$s are already introduced in
equations (1) and (7). If we define mean energies of
stellar and nebular LyC as 
$\langle \epsilon_* \rangle \equiv L_*^{\rm LyC}/Q_*$ and 
$\langle \epsilon_{\rm neb}\rangle\equiv L_{\rm neb}^{\rm LyC}/Q_{\rm neb}$, 
equation (9) is reduced to 
\begin{equation}
 \langle \epsilon_{\rm esc} \rangle 
  = \beta \langle \epsilon_* \rangle 
  + (1-\beta) \langle \epsilon_{\rm neb} \rangle\,,
\end{equation}
where 
\begin{equation}
 \beta = \left(\frac{\alpha_{\rm B}}{\alpha_{\rm A}}\right) 
  + \left(\frac{\alpha_1}{\alpha_{\rm A}}\right) f_{\rm esc}\,.
\end{equation}
Note that 
$\langle \epsilon_{\rm neb}\rangle=\gamma_{\rm neb}^{\rm LyC}/\alpha_1$, 
where $\gamma_{\rm neb}^{\rm LyC}$ is the frequency integrated emission
coefficient of nebular LyC. Therefore, the mean energy of escaping LyC
is given by the internal division of mean energies of stellar and
nebular LyC with weights $\beta$ and $1-\beta$, and thus, the escaping
mean energy is always smaller than the stellar one if $f_{\rm esc}<1$.

\begin{table}
 \caption[]{Coefficients for $\beta$ in equation (11) and mean energy of
 nebular Lyman continuum.}
 \setlength{\tabcolsep}{3pt}
 \footnotesize
 \begin{minipage}{\linewidth}
  \begin{tabular}{lcccc}
   \hline
   $T_{\rm e}$ ($10^4$ K) & $0.5$ & $1$ & $2$ & $5$ \\
   \hline
   $\alpha_{\rm B}/\alpha_{\rm A}$ & $0.669$ & $0.626$ & $0.573$ & $0.502$\\
   $\alpha_1/\alpha_{\rm A}$ & $0.331$ & $0.374$ & $0.427$ & $0.498$ \\
   \hline
   $\langle \epsilon_{\rm neb} \rangle$ (eV) & 16.41 & 16.61 & 16.98 & 19.12 \\
   \hline
  \end{tabular}
 \end{minipage}
\end{table}%

\begin{table}
 \caption[]{Mean energies of stellar and escaping Lyman continua.}
 \setlength{\tabcolsep}{3pt}
 \footnotesize
 \begin{minipage}{\linewidth}
  \begin{tabular}{lcccc}
   \hline
   Model & A & B & C & D \\
   \hline
   $\langle \epsilon_* \rangle$ (eV) 
   & 22.66 & 23.00 & 26.85 & 32.18 \\
   \hline
   $\langle \epsilon_{\rm esc} \rangle$ (eV)$^a$
   & 21.53 & 21.81 & 24.94 & 29.27\\
   $\langle \epsilon_{\rm esc} \rangle$ (eV)$^b$
   & 20.62 & 20.85 & 23.40 & 26.93\\
   \hline
  \end{tabular}

  $^a$ $T_{\rm e}=10^4$ K, $f_{\rm esc}=0.5$\\
  $^b$ $T_{\rm e}=10^4$ K, $f_{\rm esc}=0.1$\\
 \end{minipage}
\end{table}%

In Table~2 we summarize the coefficients for $\beta$ in equation (11)
and the mean photon energy of nebular LyC as a function of $T_{\rm e}$. 
Table~3 gives a summary of mean photon energies of stellar LyC for
models A--D in the first row. The second and third rows in Table~3 show
mean photon energies of escaping stellar+nebular LyC for two cases of
$T_{\rm e}=10^4$ K and $f_{\rm esc}=0.5$ or 0.1. We find that the energy
reduction is about 5--15\% for these cases and its maximum is 10--20\%
at the limit of $f_{\rm esc}\to0$.

\section{Discussion}

\subsection{Wavelength dependent opacity and comparison with the code Cloudy}

As described in \S2.2.1, LyC opacity through a clumpy ISM is independent
of wavelength. As an opposite limit, we here consider a homogeneous ISM
where the opacity keeps the wavelength dependence of the
photo-ionization cross section of hydrogen: $\propto\lambda^3$.

\begin{figure}
 \begin{center}
  \includegraphics[width=7cm]{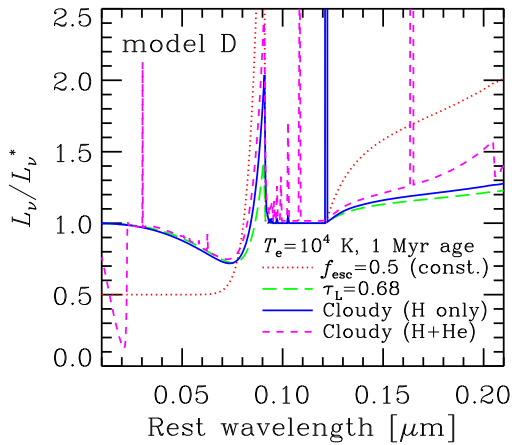}
 \end{center}
 \caption{Spectra relative to the stellar one and comparison with
 results from Cloudy. The electron temperature $T_{\rm e}=10^4$ K and
 the age of 1 Myr after an instantaneous burst for the stellar
 population model D are assumed. The dotted line is the escaping
 stellar+nebular spectrum with a constant escape fraction $f_{\rm
 esc}=0.5$ which is the same as in Figures 3 and 4. The long-dashed line
 is the escaping stellar+nebular spectra for which we assume the neutral
 hydrogen optical depth at the Lyman limit $\tau_{\rm L}=0.68$ and at
 shorter wavelengths to be $\propto\lambda^3$. The solid and
 short-dashed lines are the cases calculated with Cloudy 08.00 (Ferland
 et al.~1998) for hydrogen only case and hydrogen and helium case,
 respectively. The Lyman limit optical depth for the hydrogen only case
 is matched to 0.68.}
\end{figure}

Figure~9 shows escaping stellar+nebular spectra relative to the stellar
one in a homogeneous ISM (LyC opacity depends on the cube of wavelength)
for the model D. The Lyman limit optical depth is set to be 0.68. This
opacity corresponds to $f_{\rm esc}=0.5$ at the Lyman limit and is
applied to both stellar and nebular LyC (long-dashed line). If we
compare the case in a clumpy ISM presented in the previous sections
(dotted line), we find that the Lyman limit bump predicted in a
homogeneous ISM is weaker than that in a clumpy ISM. This is because
higher energy LyC escapes easily and does not convert to nebular
emission in the wavelength dependent case (i.e. homogeneous ISM). To
make a strong bump, we need to convert higher energy LyC to the Lyman
limit by clumpy nebulae.

We can use the public photo-ionization code, Cloudy \citep{fer98} for a
homogeneous ISM. To check our nebular calculations, we compare our
results with the code Cloudy in Figure~9. We input the following
condition to the code: a point source with the production rate of
hydrogen LyC photons of $10^{52.01}$ s$^{-1}$, which corresponds to
$10^4$ $M_\odot$ in the stellar mass for the model D, resides at the
center of spherical symmetric gas with the hydrogen atomic number density
of 1 cm$^{-3}$ and with the temperature of $10^4$ K. We set the inner
and outer boundaries to be $10^{20}$ cm and $10^{21}$ cm, respectively. 
The outer boundary is determined to achieve a Lyman limit optical depth
of 0.68. This case is shown by the solid line in Figure~9. We find a
good agreement between our calculation (long-dashed) and Cloudy
(solid). However, Cloudy predicts somewhat stronger bump at the Lyman
limit. This is because the optical depth for nebular LyC is smaller than
that for stellar one since the nebula surrounds the source, whereas we
set the same optical depth for both continua.

The short-dashed line in Figure~9 is the case where the spherical gas
consists of hydrogen and helium. Other settings are the same as the
hydrogen only case. The resultant continuum is not very different from
the hydrogen only case. Thus, the effect of helium is small for the
shape of the SED. This point is also discussed in the next subsection.

\subsection{Effect of helium}

We can derive equations for helium nebular emissions like equation (4) 
for hydrogen. The strength of the helium emissions relative to that of
hydrogen can be approximated to
\begin{equation}
 \frac{L_\nu^X}{L_\nu^{\rm H}} \approx
  \left(\frac{\gamma_\nu^X}{\gamma_\nu^{\rm H}}\right)
  \left(\frac{Q_*^X}{Q_*^{\rm H}}\right)
  \left(\frac{\alpha_{\rm B}^{\rm H}}{\alpha_{\rm B}^X}\right)\,,
\end{equation}
where $X$ is neutral helium (He I) or singly ionized helium (He II) and 
other quantities have the same meanings as in \S2 but with superscript
describing which atom or ion. In the four models of A--D considered
here, we found $Q_*^{\rm HeI}/Q_*^{\rm HI}=0.4$--0.7 and 
$Q_*^{\rm HeII}/Q_*^{\rm HI}=4\times10^{-4}$--0.1 with a larger value
for a smaller metallicity \citep{sch02,sch03}. The recombination rates
of helium are $\alpha_{\rm B}^{\rm HeI}\approx\alpha_{\rm B}^{\rm HI}$ 
and $\alpha_{\rm B}^{\rm HeII}\approx4\alpha_{\rm B}^{\rm HI}$
\citep{ost06}. According to the emission coefficients shown in
\cite{ost06}, we found that the helium contribution is negligible in the 
wavelength range of 800--2000 \AA, except for the model D where the He
II bound-free Paschen continuum contributes about 20\% around 2000 \AA.
This smallness of the effect of helium on continuum can be confirmed in
Figure~9 if we look at the result by Cloudy with hydrogen and helium
(short-dashed line). However, the effect on recombination lines may not
be negligible.

\subsection{Lyman ``bump'' as a tool to find exotic stellar populations}

As shown in Figure~4, we predict that galaxies dominated by very young
($\sim1$ Myr) and massive stellar populations can have a Lyman limit
``bump'' not ``break'' expected in the stellar spectra. The strength of
the bump depends on the hardness of the LyC (see also Figure~6). If
galaxies contain very massive ($\sim100$ $M_\odot$) EMP stars or
metal-free (Pop-III) stars, the Lyman limit bump becomes very
strong. This fact provides a new method to find such exotic stellar
populations in galaxies at $z\la4$ although the IGM attenuation hampers
us to use this method for galaxies at $z>4$. Indeed, \cite{iwa09} have
found very interesting LAEs at $z=3.1$ which emit extremely strong
LyC. We will apply the model developed in this paper to the LAEs in the
forthcoming paper and discuss their nature in detail.

\subsection{Effect on reionization}

Let us examine if escaping nebular LyC affects the reionization process.
The reionization history is regulated by the product of escape fraction
and star formation efficiency, $f_{\rm esc}f_*$, if we assume a
cosmological structure formation scenario \citep{wyi03}. As the first
effect of escaping nebular LyC, this $f_{\rm esc}$ should be replaced by
``effective'' escape fraction, $Q_{\rm esc}/Q_*$, which is introduced in
equation (8) and Figure~8. From equations (8) and (11), we obtain 
$Q_{\rm esc}/Q_*=f_{\rm esc}/\beta$ and find that the ``effective''
escape fraction is at most a factor of 1.6 larger than $f_{\rm esc}$ for
$T_{\rm e}=10^4$ K. As the second effect, the nebular LyC modifies the
LyC spectral shape and reduces the mean photon energy of escaping
LyC. This leads to a decrease of the mean free path of LyC photons in
the IGM. However, the reduction of the mean photon energy is at most
10--20\% relative to that of the stellar LyC as discussed in \S3.4 and
in Tables~2 and 3. Therefore, the nebular LyC escape itself may not give
a significant impact on the reionization process.

\section{Conclusion}

The discovery of $z=3.1$ LAEs emitting extremely strong LyC by the
Subaru telescope \citep{iwa09} requires us to examine LyC emissivity and
spectrum of galaxies more carefully. As an attempt, this paper
has examined the effect of an escape of nebular LyC on SEDs of
galaxies. The nebular LyC mainly emitted by bound-free
transitions of hydrogen is usually assumed to be absorbed within
nebulae, so-called ``on-the-spot'' approximation \citep{ost06}. However,
we should consider its escape if stellar LyC escapes from galaxies as
observed at $z\sim3$ \citep{ste01,sha06,iwa09}. 

Since the bound-free nebular emission has strong peaks at the limits of
Lyman, Balmer, Paschen, etc. (see Figure~2), we expect a bump at the
Lyman limit due to escaping nebular bound-free LyC (see Figure~3). This
bump boosts luminosity density at 900 \AA. As a result, 900 \AA-to-1500
\AA\ luminosity density ratio can become larger than that expected in
stellar SEDs. Indeed, the ratio has a maximum not when $f_{\rm esc}=1$
but when $f_{\rm esc}=0.6$ (see Figure~6). The strength of the bump
depends on $T_{\rm e}$ in nebulae (see Figure~5), hardness of stellar
LyC (i.e. metallicity, IMF and mass range, and age; see Figures~3 and
7), and IGM attenuation (see Figure~4). On the other hand, the total
number of LyC photons which escape from galaxies is always less than
that produced by stars (see Figure~8). In fact, the Lyman limit bump is
a result of the radiation energy re-distribution by nebulae. The energy
re-distribution reduces the mean energy of escaping LyC
photons, but it is at most 10--20\% reduction relative to the mean
energy of stellar LyC (see Table~3). 

We have assumed wavelength independent $f_{\rm esc}$ (or LyC opacity
through the ISM) to obtain our main results. This should be suitable for
a clumpy ISM (see schematic Figure~1 and discussion in \S2.2.1). We have
also calculated a wavelength dependent case which corresponds to a
homogeneous ISM, and have found that the wavelength dependent case
predicts a weaker Lyman limit bump (see Figure~9). In addition, we have
compared the case with the photo-ionization code Cloudy \citep{fer98},
and have found a good agreement (see Figure~9). Moreover, we have
omitted helium, but it is justified by a discussion in \S4.2 and
by a comparison with Cloudy (see Figure~9).

The most interesting implication from our results would be the
prediction of Lyman ``bump'' galaxies (see Figure~4 and \S4.3); galaxies
containing very young ($\sim1$ Myr), very massive ($\sim100$
$M_\odot$), and extremely metal-poor (or metal-free) stellar populations
show a prominent bump just below the Lyman limit. This would be a new
indicator to find such exotic stellar populations. Although IGM
attenuation hampers us to use this method for $z>5$ galaxies, we can use
it for $z<4$ galaxies.

Finally, we have discussed effects on the reionization briefly (\S4.4). 
Escaping nebular LyC has two effects: acting as an additional LyC source
and reducing the mean energy of escaping LyC photons. The first effect
increases the escape fraction effectively, but the enhancement is at
most a factor of 1.6 (for $T_{\rm e}=10^4$ K) and the LyC photon escape
rate never exceeds the production rate by stars (see Figure~8). The
second effect decreases the mean free path of LyC photons in the IGM. 
However, the reduction of the mean photon energy is at most 10--20\%
relative to the stellar LyC (see Table~3). Therefore, the nebular effect
on the reionization may be small.

\section*{Acknowledgements}

The author appreciates discussions with H.~Yajima which inspired the
idea of nebular emissions. The author would like to thank D.~Schaerer
for providing his spectral models through I.~Iwata, and C.~Leitherer and
G.~Ferland for offering their codes to public. The author is grateful to
I.~Iwata, K.~Kousai, T.~Yamada, T.~Hayashino, M.~Akiyama, Y.~Matsuda,
J.-M.~Deharveng, C.~Tapken, S.~Noll, D.~Burgarella, Y.~Nakamura, and
H.~Furusawa for discussions and comments which were very useful to
accomplish this work. The author is also grateful to T.~Kozasa and
A.~Habe in Hokkaido University for their hospitality during writing this
manuscript in Sapporo. The author is supported by KAKENHI (the
Grant-in-Aid for Young Scientists B: 19740108) by The Ministry of
Education, Culture, Sports, Science and Technology (MEXT) of Japan.

\label{lastpage}

\end{document}